\documentclass[11pt,preprint]{aastex}

\newcommand{\ea}{{et al.~}}

\newcommand{\ha}{H~$\alpha$}
\newcommand{\am}{\overline{\gamma}}
\newcommand{\dvc}{$r^{1/4}$}
\newcommand{\ser}{$r^{n}$}

\shorttitle{Nuclear properties of spirals galaxies}
\shortauthors{C. Scarlata, M. Stiavelli, M.A. Hughes et al.}

\begin{document}

\title{Nuclear properties of a sample of nearby spirals from STIS imaging
}

\author{C. Scarlata\altaffilmark{1,2}, 
M.~Stiavelli\altaffilmark{2},
M.~A.~Hughes\altaffilmark{3}, 
D.~Axon\altaffilmark{3,4}, 
A.~Alonso-Herrero\altaffilmark{5},
J.~Atkinson\altaffilmark{3}, 
D.~Batcheldor\altaffilmark{3}, 
J.~Binney\altaffilmark{6}, 
A.~Capetti\altaffilmark{7},
C.~M.~Carollo\altaffilmark{8},
L.~Dressel\altaffilmark{2},
J.~Gerssen\altaffilmark{2},
D.~Macchetto\altaffilmark{2},
W.~Maciejewski\altaffilmark{9}
A.~Marconi\altaffilmark{9}, 
M.~Merrifield\altaffilmark{10}, 
M.~Ruiz\altaffilmark{3},
W.~Sparks\altaffilmark{2},
Z.~Tsvetanov\altaffilmark{11},
R.~P.~van~der~Marel\altaffilmark{2}
}

\altaffiltext{1}{Department of Astronomy, Univerist\`a degli Studi di
Padova, Padova, Italy}
\altaffiltext{2}{Space Telescope Science Institute, 3700 San Martin
Drive, Baltimore, MD 21218}
\altaffiltext{3}{Department of Physics, Astronomy and Mathematics, University of
Hertfordshire, Hatfield, Hertfordshire, AL10 9AB, UK}
\altaffiltext{4}{Department of Physics, RIT, 84 Lomb Memorial Dr., 
Rochester, NY 14623-5603}
\altaffiltext{5}{Steward Observatory, University of Arizona, 933
North Cherry Avenue, Tucson, AZ 85721}
\altaffiltext{6}{Oxford University, Theoretical Physics, Keble Road,
Oxford, OX1 3NP, UK}
\altaffiltext{7}{INAF-Osservatorio Astronomico di Torino, I-10025 Pino
Torinese, Italy}
\altaffiltext{8}{Physics Department, ETH, Hoenggerberg HPF G4.3, 
CH--8092 Zurich, Switzerland}
\altaffiltext{9}{INAF-Osservatorio Astrofisico di Arcetri, Largo E. Fermi
5, 50125 Firenze, Italy}
\altaffiltext{10}{School of Physics and Astronomy, University of
Nottingham, NG7 2RD, UK}
\altaffiltext{11}{Center for Astrophysical Sciences, Johns Hopkins
University, 239 Bloomberg Center for Physics \& Astronomy, 3400 North
Charles Street, Baltimore, MD 21218}

\begin{abstract}
We present surface photometry for the central regions of a sample of
48 spiral galaxies (mostly unbarred and barred of types Sbc or Sc)
observed with the Space Telescope Imaging Spectrograph on board the
Hubble Space Telescope. Surface brightness profiles were derived and
modeled with a Nuker law. We also analyzed archival Wide Field
Planetary Camera~2 images with a larger field of view, available for
18 galaxies in our sample. We modeled the extracted bulge surface
brightness profiles with an exponential, an \dvc, or an \ser\
profile. In agreement with previous studies, we find that bulges of
Sbc galaxies fall into two categories: bulges well described by an
exponential profile and those well described by an \dvc\ profile. Only
one galaxy requires the use of a more general Sersic profile to
properly describe the bulge. Nuclear photometrically distinct
components are found in $\sim$ 55\% of the galaxies. For those that we
classify as star clusters based on their resolved extent we find
absolute magnitudes that are brighter on average than those previously
identified in spiral galaxies. This might be due to a bias in our
sample toward star forming galaxies, combined with a trend for star
forming galaxies to host brighter central clusters.

\end{abstract}


\section{Introduction}
This article is part of a series of papers presenting the results of
our Hubble Space Telescope (HST) program GO--8228 (PI: D. Axon),
executed with the Space Telescope Imaging Spectrograph (STIS). The
goal of the program is to study the black hole (hereafter BH) mass
distribution in spiral galaxies.

Studies of the centers of nearby early--type galaxies (ellipticals and
lenticulars) have revealed that most contain supermassive black holes
(see \citealt{kormendy3} for a recent review). These studies also
revealed a strong correlation between the mass of the BH ($M_{\rm
BH}$) and the mass (or luminosity) of the host spheroid ($M_{\rm
sph}$, \citealt{magorrian1}; \citealt{marconi2}).  An even tighter
correlation was thought to exist between $M_{\rm BH}$ and the mean
velocity dispersion ($\sigma$) of the bulge, measured inside its
effective radius (\citealt{ferrarese2}; \citealt{gebhardt1}). However,
\citet{marconi2} have recently shown that when considering only
galaxies with secure BH detections the above correlations have similar
intrinsic dispersion. In contrast, $M_{\rm BH}$ is unrelated to the
properties of galaxy disks. These correlations strongly support the
idea that the growth of BHs and the formation of spheroids are closely
linked (\citealt{silk1}; \citealt{haehnelt1}). However, these
results are based on samples strongly biased against late Hubble type
galaxies: only $\simeq 20$\% of the detected BHs are in spiral
galaxies (\citealt{kormendy3}; \citealt{merritt1}).

There are significant differences between bulges of intermediate and
late--type galaxies and those of early--type spirals
(\citealt{kormendy1}; \citealt{carollo1}; Carollo, Stiavelli, \& Mack
1998). The formers (often referred as ``pseudobulges'') are
characterized by exponential surface brightness profiles, shallow
nuclear cusp slope (\citealt{carollo3}), and cold kinematics
(\citealt{kormendy1}). In contrast, spheroids of early type spirals
are better described by a de Vaucouleurs \dvc\ profile, have steep
stellar cusps that increase with decreasing luminosity
(\citealt{carollo3}), and are, in general more similar to small
ellipticals than to exponential bulges. \citet{carollo4} suggested
that exponential and \dvc\ bulges have different origins. While it
seems reasonable (in light of the observational results) that the
former form within the disk as a consequence of the secular evolution
of the disk itself (\citealt{norman1}), it is difficult to explain the
origin of the \dvc\ bulges with the same mechanism. Therefore, there
is the possibility that the $M_{\rm BH}/M_{\rm sph}$ relation for
spirals could be different for the two classes of bulges, mirroring
the different formation processes. In order to address this, it is
necessary to have access to high spatial resolution spectroscopy (to
measure the BH mass) and imaging (to characterize the nuclear surface
brightness profile). The observing program we are conducting is ideal
for these kinds of studies.  In fact, the BH mass will be derived by
modeling the gas kinematics, and the nuclear properties of the bulges
will be derived by analyzing the high spatial resolution images taken
to accurately center the galaxy nucleus in the slit.

In this paper we present the STIS images obtained during our HST
program. We use these data, together with archival HST--Wide Field
Planetary Camera~2 data available for some of the galaxies, to
describe the nuclear and bulge properties of the galaxies in the
sample.  The paper is organized as follows: the data, the data
reduction, and the extraction of the radial surface brightness
profiles are described in Section~\ref{sec:data}. In
Section~\ref{sec:modeling} we discuss the radial profile fits used to
derive the nuclear slope and bulge type. The principal results
concerning the nuclear slopes, and galaxy types are presented and
discussed in Section~\ref{sec:results}. In Section~\ref{sec:nuclei} we
discuss the identification of nuclear star clusters, and we derive
their main properties such as size and total magnitude. Lastly, in
Section~\ref{sec:conclusion} we discuss the results.

In paper I (\citealt{marconi1}) we presented a detailed description of
the modelling techniques used to determine the BH mass from the STIS
data. The model was then applied to the images and spectra for NGC
4041, the first galaxy to be observed in the HST program. In Paper II
(\citealt{hughes1}) we presented the STIS spectra, most of the STIS
images, and color maps when archival near--infrared NICMOS images were
available. In Paper IV (Hughes et al., in prep) we use both color
information and the spectra to investigate the ages of the central
stellar population.  The present paper is Paper III in the series.

\section{Observations and Data Reduction} 
\label{sec:data}

The galaxy sample observed with STIS consists of 54 spiral galaxies,
mostly classified as Sbc and Sc. The galaxies were extracted from a
larger sample (128 objects) for which we obtained ground--based \ha\
rotation curves at a seeing--limited resolution of
$1^{\prime\prime}$. VLA radio maps and ground--based $R$ and $B-$band
CCD images are also available for all the galaxies. The 54 galaxies
were chosen to have recession velocity $< 2000$ km s$^{-1}$ (in order
to be able to resolve the rotation curves within the central few
parsecs), and to show emission lines in the nucleus.

The sample galaxies together with their main properties are listed in
Table~\ref{tab:sample}. Columns~2 and 3 list the morphological
classification and the apparent total $B$ magnitude from the Third
Reference Catalog of Bright Galaxies (RC3;
\citet{devaucouleurs1}). The galactic $R-$band extinction, the
distance, and the physical pixel scale are presented in Columns~4--6,
respectively.  The last column of Table~\ref{tab:sample} gives a brief
description of the nuclear morphology of the galaxies. Throughout this
paper we adopt H$_0=65$ km sec$^{-1}$ Mpc$^{-1}$.

\subsection{STIS Data}

The STIS images were taken as part of the acquisition procedure to
accurately center the STIS slit on the galaxy nuclei. They were
obtained between July 1999 and February 2001. We adopted the STIS
diffuse--source acquisition procedure (see the STIS Instrument
Handbook for details, \citet{proffitt1}). During the acquisition two
$5'' \times 5''$ images were obtained with the optical long--pass
filter F28X50LP. After the first image was created from a pair of
exposures (see chapter 8 of the STIS Instrument Handbook for details),
the position of the target was computed by finding the flux--weighted
centroid of the pixels in the brightest checkbox (7 pixels size) and a
second exposure was made from a second pair of exposures with the
galaxy nucleus centered in the acquisition aperture. For NGC~134,
NGC~3521, NGC~3972, NGC~4389, NGC~5577 and NGC~5713, the slit was not
correctly placed on the nucleus of the galaxy. These galaxies are
indicated with the word ``missed'' in Table~\ref{tab:sample}. In these
cases, either the centering algorithm did not work properly, or our
coordinates of the targets were insufficiently accurate. The pixel
scale of the images is 0\farcs05 pixel$^{-1}$.

All the images were reduced using the most updated reference
files. The raw images were flat--field corrected using the task
BASIC2D in the package STSDAS in IRAF\footnote{IRAF is distributed by
the National Optical Astronomy Observatories, which are operated by
AURA, Inc., under cooperative agreement with the National Science
Foundation.}. The processing done by the flight software is
rudimentary, in particular a single predefined bias value of 1510 DN
is subtracted. This value is an approximation of the actual bias,
which has been observed to increase with time (by the end of 2000 the
actual level in the acquisition subarray was $1517.3 \pm 1.7$ DN). 
Since our observations cover a period of time that goes from July 1999
to February 2001, it is necessary to correct for this effect. We have
therefore retrieved from the HST archive all bias images taken as part
of the STIS CCD Performance Monitor program, from June 1998 to May
2001. The bias level variation as function of time was fitted with a
linear function, and each galaxy image was corrected accordingly.

The two images of the target were aligned and then averaged with
simultaneous rejection of residual cosmic rays not removed by the
onboard processing of the pairs of exposures. The shifts were computed
from the position of the center of the galaxy and any star clusters or
point sources visible in the two exposures.  For two objects (namely
NGC~3003, NGC~4088) only the second exposure contained useful
data. For these objects the analysis was performed on the single
second exposure. Residual cosmic ray rejection, for these exposures,
was performed with the IRAF task CRMEDIAN.

In order to facilitate the comparison between our results and those of
other studies, we calibrated the F28X50LP magnitudes to Cousins
$R$. The photometric zero point was computed using the package SYNPHOT
in STSDAS, using different galaxy templates (\citealt{kinney1}). We
found that the zero points derived for spiral galaxy templates (Sb and
Sc templates) do not show significant differences from each other
($< 2$\%), while the zero point computed for the elliptical
galaxy template differed more than 5\% from the others. Since we do
not have color information for the nuclear region of the observed
galaxies, and since the entire sample mostly consists of Sbc type
galaxies, we decided to adopt the calibration computed for the Sb
template. All the magnitudes were expressed in VEGAMAG (the spectrum
of the star Vega is used as zero point), and were corrected for
Galactic extinction following Schlegel, Finkbeiner \& Davis (1988). In
order to convert from VEGAMAG to ABMAG (a magnitude system with zero
point based on a spectrum with constant flux per unit frequency, Oke
1974), the transformation for an Sb galaxy spectral template is:
$m_{R,{\rm Vega}}=m_{R,{\rm AB}}-0.21$. In Figure~\ref{fig:images1} we
present the reduced images of the galaxies. The nucleus is not always
at the center of the field because the images are a mosaic of two
exposures.

\subsection{WFPC2 Data}
The field of view of the STIS images is only $5'' \times 5''$ which in
physical units translates to between $0.2\times0.2$ kpc and
$0.7\times0.7$ kpc, depending on the distance of the galaxy. Although
this provides important information about the structure of the
innermost region, it does not fully cover the bulge and the disk
component of the galaxy. For this reason we searched the HST archive
for WFPC2 images of the sample galaxies. We found data for 18
objects. Most of the galaxies were observed with the filter F606W,
three galaxies (NGC~5248, NGC~7314, and NGC~7331) with F814W, and two
(NGC~4321 and NGC~4536) with F555W. We retrieved from the HST archive
the on--the--fly calibrated images that are reprocessed with the most
recent reference frames for flat-fielding, bias, and dark current
subtraction.  The different exposures of the same target were combined
and cosmic--ray cleaned in a single step performed using the
IRAF/STSDAS task CRREJ. Absolute photometric calibration was obtained
by using the photometric zero points for the filters provided by
\citet{holtzman1} to facilitate the comparison with the results of
\citet{carollo3}. The magnitude in the F814W filter was
then used to estimate the $V-$band magnitude using SYNPHOT in
STSDAS. Since this correction depends on the spectral energy
distribution of the object, it was calculated using the
\citet{kinney1} Sb spectral template. The magnitudes were corrected
for galactic extinction following \citet{schlegel1}.

\subsection{Surface brightness profiles}
For each galaxy we extracted the surface brightness profile (SBP) by
fitting ellipses to the isophotes using the isophote--fitting program
ELLIPSE in IRAF.  Before running ELLIPSE on the images we corrected
for the presence of dust lanes, following the technique introduced by
\citet{carollo1}.  Briefly, a 2--dimensional model of the galaxy
was created assuming constant ellipticity and PA with radius, equal
to the values for the outermost isophotes.  The difference between the
model and the original data gives an image with flux only in the
regions of dust absorption. Pixels with significant flux in the
difference image (above 3 $\sigma$ level) were replaced on the actual
image by the model. This procedure was repeated iteratively until
convergence was reached. Typically only two or three iterations were
needed.  

This technique allows us to correct for patchy/filamentary dust
absorption and has the advantage of being independent of any
physical model of the nature of the dust and its distribution with
respect to the stars. It does not correct for any smooth or
extended dust component. However, the influence of an
extended/diffuse dust component on the structural parameters
derived from the SBP fit was studied in detail by Carollo (1999)
using optical and infrared data of the nuclear regions of a sample
of spiral galaxies. This work showed that the photometric
structural parameters derived from optical images, when corrected
with the Carollo et al. (1997) algorithm, are accurate and do not
suffer significantly from any residual uncorrected dust extinction.

The adopted method can be used to derive an estimate of the extinction
in each pixel of the image. Following \citet{witt1992} we defined the
effective optical depth (absorption $+$ scattering: $\tau_{\rm eff}$)
as $\tau_{\rm eff}= -\ln{I_{(\rm obs}/I_{mod})}$, where $I_{\rm obs}$
is the observed flux density and $I_{\rm mod}$ is the flux density in
the reconstructed 2--dimensional model. For most of the galaxies we
found that $\simeq 85$\% of the pixels to which a correction was
applied have $\tau_{\rm eff}<0.4$. This corresponds to $\tau_{\rm
eff,V}< 0.6$. The conversion between $\tau_{\rm eff}$ (which is the
average over the wavelengths included in the F28X50LP filter) and
$\tau_{\rm eff,V}$ was computed by using the spectral template of an
Sb galaxy, and using the extinction law as given in
\citet{cardelli1989}. For these values of optical depth the
contribution of scattered light to the total emerging light is less
than 5\%, unless particular geometries in the dust/stars distribution
are considered.

The extracted $R-$band STIS surface brightness profiles for 40
galaxies are presented in Figure~\ref{fig:profiles1}. For a few of the
sample galaxies, the patches and lanes of dust are too prominent
(e.g., NGC~4527) to allow a meaningful correction for its effect on
the galaxy surface brightness profile with the Carollo et al.~(1997)
algorithm. In a few other galaxies either the presence of bright knots
of star formation or a too small gradient in the surface brightness
made it impossible to obtain a good description of the galaxy with an
elliptical isophote analysis. For these reasons we failed to obtain a
meaningful isophotal fit for 8 galaxies in the STIS sample, and 3 in
the WFPC2 sample. In Figure~\ref{fig:wfpcprof} we present the $V-$band
radial surface brightness profiles measured on the WFPC images.

\section{Surface brightness profile modeling}
\label{sec:modeling}
\subsection{Average logarithmic slope $\am$}
Since the radial extent of the extracted STIS SBPs is only a few
arcseconds, we decided to fit them with a ``Nuker law''
(\citealt{lauer1}), without trying any decomposition into bulge/disk
component in the STIS data.  The Nuker law has been proved to
accurately describe the central part of galaxy profiles both for
early--type galaxies (\citealt{ferrarese1}; \citet{forbes1};
\citealt{lauer1}, \citealt{ravindranath1}; \citealt{laine1}) and spiral
galaxies (\citealt{carollo1}; \citealt{seigar1}). It assumes that the
SBP is a combination of two power laws with different slopes
($\gamma$) and ($\beta$) for the inner and the outer regions. The
radius at which the transition between the two power laws occurs is
the break radius $r_b$. $I_b$ is the surface brightness at $r_b$, and
the sharpness of the transition is described by the parameter
$\alpha$. The functional form of the SBP is:

\begin{equation}
I(r)=2^{(\beta-\gamma)/\alpha}I_{b}\left(\frac{r}{r_b}\right)^{-\gamma}\left[1+\left(\frac{r}{r_b}\right)^{\alpha}\right]^{(\gamma-\beta)/\alpha}.
\end{equation}

In order to derive the parameters describing the galaxy profiles
[$\alpha$, $\beta$, $\gamma$, $I_b$, and $r_b$], we iteratively fitted
the SBP model to the observations using a non--linear $\chi^2$
minimization based on the Levenberg--Marquardt method (e.g.,
\citealt{bevington1}). The fit was done over a user--specified radial
range, taking into account the instrumental Point Spread Function
(PSF) by convolving the model profile with the appropriate PSF. The
latter was created for the STIS--F28X50LP setup using the Tiny Tim
software (\citealt{krist1}).  The convolution was performed in the
two--dimensional plane of the sky as a product in the Fourier domain,
before the $\chi^2$ minimization.  In order to avoid numerical
artifacts, the model and the PSF were super--sampled by a factor of 5
before the convolution.

The presence of a photometrically distinct nuclear component is one of
the main sources of uncertainty in deriving the parameters describing
the underling galaxy light (e.g., \citealt{carollo2}). One possible
approach is to fit a two--component model to the SBP, with one
component describing the nuclear--source profile and the other the
underlying galaxy (\citealt{boeker2}). Another possibility is to fit
the model describing the galaxy only in a region not affected by the
nuclear source (\citealt{stiavelli}). We decided to adopt the second
strategy, since we were mainly interested in measuring the central
slope of the galaxy SBP.  Furthermore, it is not clear which
analytical form is convenient to use to describe these nuclear
components, since they can be dominated by AGN light (point--source
like profile), by a star cluster or by a combination of the two.

The radius $r_0$ starting from which the SBP was considered unaffected
by light coming from the central component was identified by eye
for each galaxy's SBP. We typically found $r_0$ ranging from
0\farcs3 to 0\farcs5.  Since the choice of $r_0$ is somewhat subjective
(being the extent of the central component estimated visually from
the profile), it can be a source of errors. We verified that
changing the radial range by $\pm$ 0\farcs05 does not significantly
alter the values of the Nuker parameters derived in the fit.

The slope of the SBPs for $r\rightarrow 0$ was computed as the mean
slope ($\am$ ) between 0\farcs1 and 0\farcs5 of the best--fit model
(\citealt{stiavelli}). Since $\am$\ is computed using the model, it
is not affected by either the STIS PSF or light coming from a nuclear
component. To verify the latter, we checked if there was any
correlation between the presence of a nuclear source (either resolved
or point--source) and the value of $\am$\ measured from the model. We
divided all galaxies in three classes: no central component,
class$=1$; point--source, class$=2$; and resolved component,
class$=3$. In Figure~\ref{fig:gamma_cluster} we plot the measured
values of $\am$\ as a function of the object class. It can be seen
that there is no correlation between the slope and the presence of a
central source, which provides confidence that the adopted strategy to
derive the parameters is correct. For $\sim 80$\% of the galaxies, the
average of the nuclear slope is in agreement with the fitted
$\gamma$. For the other objects, the fitted value is $\simeq 0$, but
this value is not reached at observationally accessible radii, given
the accompanying low values of $r_b$.  Therefore, $\am$\ is a more
robust description of the galaxy SBP. 

The fitted values of the Nuker parameters ($\alpha$, $\beta$,
$\gamma$, $r_b$, $I_b$) are listed in Table~\ref{tab:results},
together with the values of the nuclear cusp slope $\am$.  The
best--fit profiles (PSF--convolved) are overplotted on the data in
Figure~\ref{fig:profiles1}.

Nine galaxies of our sample are also present in the sample of
\citet{carollo6} for which \citet{seigar1} obtained the average
logarithmic slope of the NICMOS ($H$ band) SBPs, using the same
definition used here.  They noticed that $\am$\ derived from the
visual passband and from the IR data are in good agreement. Hence, it
is meaningful to compare our results with theirs. In
Figure~\ref{fig:seigar} we show the comparison between the average
logarithmic slope computed by Seigar \ea (2002, $\am_{H}$) from the
NICMOS images and the value ($\am_{R}$) computed here for six
galaxies. The results agree within the uncertainties.  The galaxy
(NGC~4536) for which $\am_R$ and $\am_H$ are significantly different
has a morphology which is strongly affected by dust extinction in the
STIS images (see Figure~\ref{fig:images1}). For three other galaxies
in common to our sample (NGC~2784, NGC~2903, and NGC~4527) it was not
possible to measure the SBP from our STIS data.

In principle there need not always be exact agreement between
$\overline{\gamma}_R$ and $\overline{\gamma}_H$ if there were a strong
color gradient in the nuclear regions of these galaxies due to either
a change in age or metallicity of the stellar population or if there
were uncorrected residual effects of dust extinction
\citep{witt1992}. However, the fact that a good correlation is
observed \citep{seigar1} suggests that these issues are generally not
affecting the analysis.

\subsection{Analytic fits to the bulge component}
We tried to model the more extended WFPC2 SBPs with a variety of
analytical profiles. Besides the bulge component, we also took into
account the presence of the disk of the galaxy when visible. We used
the exponential law to describe the radial surface brightness profile
of the disk component (\citealt{freeman1}). For the bulge component we
used ({\it i}) single exponential; ({\it ii}) de Vaucouleurs (\dvc);
and ({\it iii}) Sersic profile (\ser, Sersic 1968). The program used
to fit the profile was similar to the one used to fit the Nuker
law. The WFPC2 PSF was created for the different setups and different
positions on the chip that were used, using the Tiny Tim software
(\citealt{krist1}). It was convolved with the model before the fit. To
determine the nature of the bulge, we fitted the profiles only for
radii $> 1''$. This ensures that the two regions used to measure the
nuclear cusp slopes and the bulge properties are independent. In
Table~\ref{tab:results} we report the bulge classification, the scale
radius $r_{0,\rm bulge}$, the total apparent magnitude $m_{V,{\rm
bulge}}$, and the Sersic index $n$. The best--fit profiles
(PSF--convolved) are overplotted on the data in
Figure~\ref{fig:wfpcprof}.

\subsection{Results}
\label{sec:results}

We were able to measure the Nuker parameters of the nuclear profile
for 37 galaxies, out of the 40 for which we could extract the SBP.  We
find that $\simeq 56$\% of the galaxies have ``steep'' cusps ($\am
>0.5$).  The same fraction was found by \citet{carollo3} in
their total sample of 41 spiral galaxies. They found also that going
from early--type (S0a--Sa) to late--type spirals (Sc--Sd) the fraction
of \dvc\ bulges and the average nuclear slope decrease. Almost all the
galaxies in our sample are classified as Sbc in the RC3 catalog (de
\citealt{devaucouleurs1}, see Table~\ref{tab:sample}). From Figure~5 of
\citet{carollo2} we deduce that in their sample $\simeq 25\%$ of Sbc
galaxies have $\am$\ $> 0.5$. For our sample (excluding the 4 Sc
galaxies) we find that $\simeq 60$ \% of the Sbc galaxies have $\am$\
$> 0.5$.  This difference could be due to the smaller sample of Carollo
\ea (they have only 8 Sbc galaxies, out of 41 spirals) compared to our
sample (33 Sbc spirals). Furthermore, the difference between two
contiguous morphological classes is somewhat arbitrary and depends on
visual inspection of photographic sources. Considering together Hubble
types from Sb to Sc in the Carollo \ea sample, the fraction of
galaxies with ``steep'' cusps increases slightly ($\simeq 33$ \%) but
is still lower than in our sample. Our fraction is more similar to the
fraction of ``steep'' bulges in the classes S0a--Sab ($\simeq 70$\%)
in the Carollo \ea sample.

Concerning the bulge properties, we find that 8 galaxies host a bulge
that is well described by an exponential profile and 6 are well
described by an \dvc\ profile. We find only one galaxy (NGC~3887) for
which the bulge is best described by a more general \ser\ profile. All
the others have Sersic index $n$ consistent either with a pure
exponential ($n=1$) or with a pure \dvc\ ($n=0.25$) profile.

In Figure~\ref{fig:gammabulge} we show the dependence of $\am$\ on the
absolute total $V-$band magnitude of the bulge (M$_V$) for the
galaxies in our sample for which we found WFPC2 data and for which a
good fit to the bulge component could be performed ({\it filled
symbols}). We also show the objects studied in Carollo \& Stiavelli
(1998, {\it open symbols}). {\it Squares} represent bulges with an
exponential profile, while {\it circles} represent the $r^{1/4}$
bulges.  As previously shown in Carollo \ea (1998) we find that the
two classes are well divided in the $\am$\--M$_V$ plane. The only
exception is the galaxy NGC~289 for which the bulge is well described
by an exponential profile, yet it falls in the region of the \dvc\
bulges. The results about this object are uncertain since its
morphology is strongly affected by dust extinction (the most extincted
parts of the galaxy have $\tau_{\rm eff} \simeq 1$); furthermore, while
we do not identify any distinct component in the nucleus of this
galaxy, \citet{carollo6} suggest that a nuclear source is present.  It
is worth noting that the nuclear component found in \citet{carollo6}
is defined as an uncertain detection, due to a very complex central
structure.

\section{Photometrically Distinct Nuclei}
\label{sec:nuclei}
\subsection{Identification and Modeling}
Central, photometrically distinct, nuclear components either resolved
or not resolved, are common in galaxy nuclei (\citet{carollo6};
\citealt{boeker2}). The presence of such a component in a galaxy can
often be identified by looking at the image of the galaxy itself.  As
an example, NGC~4041 clearly shows a bright nucleus well distinct from
the galaxy disk (see Figure~\ref{fig:images1} and \citealt{marconi1}).
By looking at the surface brightness profile, these central components
appear as an excess of light in the central region with respect to the
best fit model computed for each galaxy. This excess can be either of
stellar nature (i.e., a nuclear stellar cluster), or non-stellar
nature (i.e., an active nucleus). In the latter case the central
component is a point source, and appears unresolved in the HST
images. Examples of these two cases are NGC~4041 where a star cluster
is present in the nucleus, and NGC~4051 which has a strong central
unresolved source (see Fig~8 in \citealt{marconi1}).

By inspecting the surface brightness profiles we identified 26
galaxies with excess light in the inner parts. Each galaxy profile was
compared with the STIS PSF profile to check whether or not the nucleus
was resolved. The results are shown in Table~\ref{tab:clusters}, where
in the second column ``PS'' indicates a point source and ``R''
indicates a resolved component.  It is difficult to address the origin
of the nuclear component (i.e., whether or not the light is dominated
by an AGN or by stellar sources) without color and spectroscopic
information.  We therefore decided to be conservative and classified
as star clusters only those nuclei that were well resolved at the
resolution of the STIS images. With this definition the identified
number of stellar clusters is a lower limit, since the appearance of a
cluster (i.e., whether or not it is resolved) strongly depends on the
distance of the galaxy and the instrumental resolution. We define a
source to be resolved when the measured Full Width at Half Maximum
(FWHM) is $\rm{FWHM} \ge 1.5 \times \rm{FWHM}_{\rm
PS}$\footnote{FWHM$_{\rm PS}$ is the FWHM of a point source}. Given
the small field of view of the STIS images we do not always have stars
suitable to measure the point source FWHM. We therefore used the STIS
PSF created with Tiny Tim and we measure FWHM$_{\rm PS}=$0\farcs08.
Given the range of distances covered by the sample, the minimum radius
that can be resolved ranges from 3 pc for the closest galaxies up to 6
pc for the most distant ones. Spectroscopic investigations of resolved
nuclear sources in spiral galaxies confirm that they are generally
star clusters (\citealt{boeker1}; \citealt{walcher1}).

The total cluster luminosity was derived using two different
techniques. In the first approach we fitted the cluster with a
Gaussian function. The contribution from the underlying bulge was
subtracted as a constant value given by the median galaxy flux within
an annulus centered on the galaxy nucleus, with radius of
$2\times\rm{FWHM}$, and thickness of 3 pixels. The total magnitude of
the cluster was then obtained by the total flux under the Gaussian
after the background subtraction. In the second approach, the galaxy
contribution below the cluster was estimated by integrating the
surface brightness of the extrapolation of the Nuker law over radii
$<r_0$ (see Section~\ref{sec:modeling}). This two methods give a lower
and upper limit to the cluster luminosity and therefore allow us to
bracket the true cluster magnitude.

The half--light radii of the clusters were estimated from the measured
Gaussian FWHM values, assuming that the intrinsic cluster profile is
well described by a Plummer law:

\begin{equation}
I_P(r)=\frac{L}{\pi\,b^2}\left(1+\frac{r^2}{b^2}\right)^{-2}.
\end{equation}

\noindent
Here $L$ is the total luminosity and $b$ is the half--light radius.
In order to correct for the STIS instrumental width, we derived the
relation between the intrinsic $b$ and the observed FWHM, by
simulating 100 cluster images (with known $b$), then convolving them
with the STIS PSF, and measuring the FWHM in the same way as done for
the actual data. Tests performed to check the accuracy of these
measurements showed that radii estimated in this way are accurate to
within 40\% (\citet{carollo6}, hereafter C02) for those clusters that
we consider resolved ($\rm{FWHM} \ge 1.5 \times \rm{FWHM}_{\rm PS}$).

The absolute total magnitudes and the corrected radii in pc are
presented in Table~\ref{tab:clusters}, columns~3--4. The listed
magnitudes are an average of the two estimates for the galaxies for
which the Nuker fit was available to estimate the background, and the
magnitudes derived with the annulus background estimate for the
others. The listed uncertainties are given by the semi--difference of
the results with the two methods.  When only the first method could be
used because no Nuker fit was available for the galaxy profile, the
errors were estimated by the semi difference of the magnitudes
obtained by varying the annulus used to estimate the underlying galaxy
contribution by $\pm$ 0\farcs05.

\subsection{Results}

We find 26 photometrically distinct components in the nuclei of our
galaxies.  This gives a fraction of $\simeq 55$\% which is comparable
with the fraction of $\simeq 50$ \% found by \citet{carollo1} in a
sample of 35 spiral galaxies, but smaller than the fraction of 75\%
found by \citet{boeker2}. However, the latter sample is mainly
composed of very late type spirals (Scd or later) where the detection
of nuclear clusters is made easier by the faintness (if not absence;
\citealt{boeker3}) of the bulge component.

We find that $\simeq 40$\% of the 26 identified nuclear sources are
not resolved; while in \citet{carollo1} most of the sources are
resolved. This difference is probably due to the different selection
criteria used to define the galaxy samples. Our sample is composed of
galaxies known to have emitting gas in the nuclear regions, therefore
possibly being biased to contain more active nuclei than Carollo's
sample.  We do not find, as in \citet{carollo2}, that {\it all} the
exponential bulges host a nuclear component, although one of the two
cases in which the source is not detected is ambiguous: NGC~289 has a
central morphology clearly affected by dust, and shows a resolved
nucleus in the infrared images studied by C02.

C02 present a correlation between the logarithm of the cluster size
and its $V-$band absolute magnitude.  They find that this correlation
exists also in the $H-$band data where the extinction by dust is less
important than in the STIS data \citep[${\rm
A_H/A_R}=0.25$,][]{cardelli1989}. For our sample of clusters we find
the same correlation with a linear correlation coefficient of $-0.52$
(significance level $> 95$\%). However the correlation disappears when
we properly account for the mutual dependence of the two variables on
distance. Indeed, using the partial correlation analysis
(\citealt{fisher1}) we find that the significance level of the
correlation between the magnitude and the radius decreases to 62.16\%
(while still significant for the C02 sample). This phenomenon is
probably due to the small size of our sample (15 objects, compared to
38 identified by C02 in the $F606W$). However, it can also be due to
the fact that we are sampling a smaller range in distances. The
comparison with the C02 sample of nuclear clusters is shown in
Fig~\ref{fig:clusters}.  Nuclear star clusters identified in this work
are represented as {\it filled circles} and {\it triangles}, while the
C02 sample is shown with {\it open circles}). {\it Triangles}
represent galaxies for which we do not have the Nuker fit of the bulge
component. Our sample seems to have brighter magnitudes for the same
cluster size, the discrepancy being larger for smaller radii. This can
be due to the fact that almost all galaxy nuclei hosting clusters are
spectroscopically classified as star forming (HII region type
spectrum) from ground--based data (\citealt{ho1}). Indeed,
\citet{carollo1} already noticed that the central sources in star
forming galaxies are typically brighter than those in non
star--forming galaxies, for similar radii.

\section{Discussion and Conclusions}
\label{sec:conclusion}
We analyzed optical images of the central regions of 48 nearby spiral
galaxies (44 Sbc, 4 Sc). The extracted SBPs were modeled with a Nuker
law, and the average logarithmic slope was computed for each best--fit
model. For a subsample of 15 objects we extracted the SBP from WFPC2
images and we derived the bulge properties by fitting an exponential,
an \dvc, or a Sersic profile.

In agreement with earlier studies, we find that galaxy bulges divide
in two classes, based on their $\am$\ and bulge SBP. In our sample of 15 Sbc
spirals we find that $\sim 50$\% of bulges are exponential, and the
other half are \dvc. This fraction is higher than the fraction of
\dvc\ bulges found by \citet{carollo3} and is more
consistent with the fraction of \dvc\ bulges found in S0--Sab
galaxies.  We also tried to fit the bulge SBP with a \ser\ law, but
the fitted $n$ values were generally found to be consistent with a
pure exponential ($n=1$) or pure \dvc ($n=0.25$) profile, implying no
need for using a more general Sersic law. Only one galaxy is best
fitted with a Sersic profile, with index $n=0.5$. This result
contrasts with a recent study by \citet{balcells1} of 19 spiral
galaxies (from S0 to Sbc morphological type), where they find no
galaxies with an \dvc\ bulge. From their Figure~2, we deduce that they
fit all the Sbc galaxies with a pure exponential profile, but their
sample contains only 4 Sbc spirals. Therefore any conclusion about the
lack of \dvc bulges in this morphological class is doubtful.

We find that $\sim 55$\% of our galaxies have a nuclear component, and
about half of them are resolved. Due to the lack of color and spectral
information, it is difficult to address the origin of the nuclei. We
conservatively decided to classify as star clusters (rather than AGN)
only those nuclei that were well resolved at the HST resolution. We
found 15 nuclear clusters, with absolute magnitudes ranging from $-10$
up to $-14.5$. The identified clusters are, on average, brighter than
those previously found. This might be due to a bias in our sample
toward star forming galaxies, combined with a trend for star forming
nuclei to host brighter central clusters.  

\bigskip
This work was funded by NASA grants for program GO-08228 from the
Space Telescope Science Institute (operated by the Association of
Universities for Research in Astronomy, Inc., under NASA contract NAS
5--26555). We made use of the NASA/IPAC Extragalactic Database, which
is operated by the Jet Propulsion Laboratory, California Institute of
Technology, under contract with NASA.

\newpage

\newpage

\begin{figure}
\caption{F28X50LP images of the galaxies in the sample. The field of view is $\simeq 5'' \times 5''$.  
\label{fig:images1}}
\end{figure}
\begin{figure}
\figurenum{1}
\caption{Continued}
\end{figure}
\begin{figure}
\figurenum{1}
\caption{Continued}
\end{figure}
\begin{figure}
\plotone{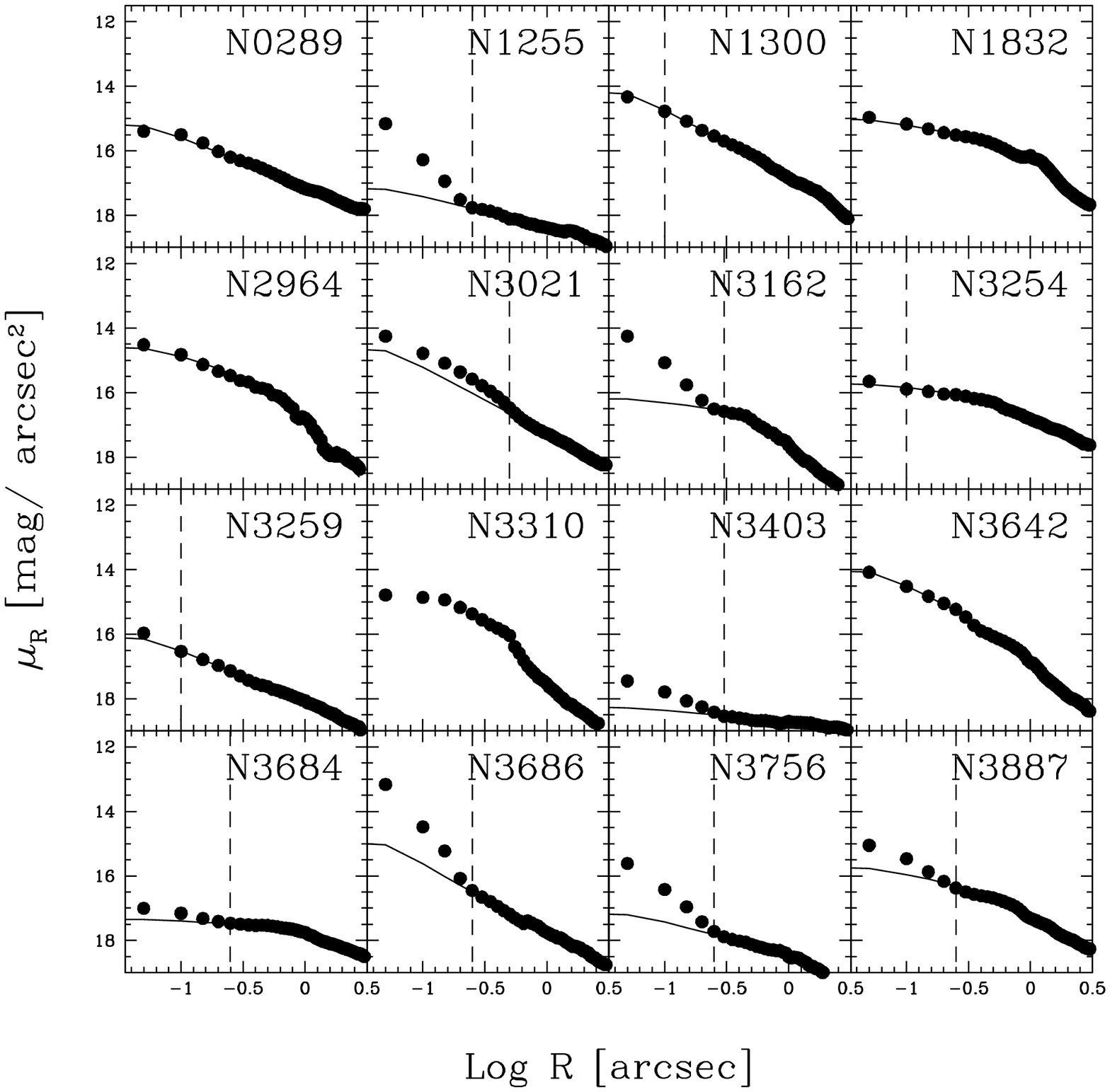}
\caption{$R-$band surface brightness profiles extracted from the STIS data ({\it dots}) and best--fit Nuker--law models ({\it solid lines}). The model was convolved with the STIS PSF before plotting. The error bars in the measured galaxy profiles are comparable with the symbol size. {\it Dashed vertical lines} indicate the inner radius used in the fit.
\label{fig:profiles1}}
\end{figure}
\begin{figure}
\figurenum{2}
\plotone{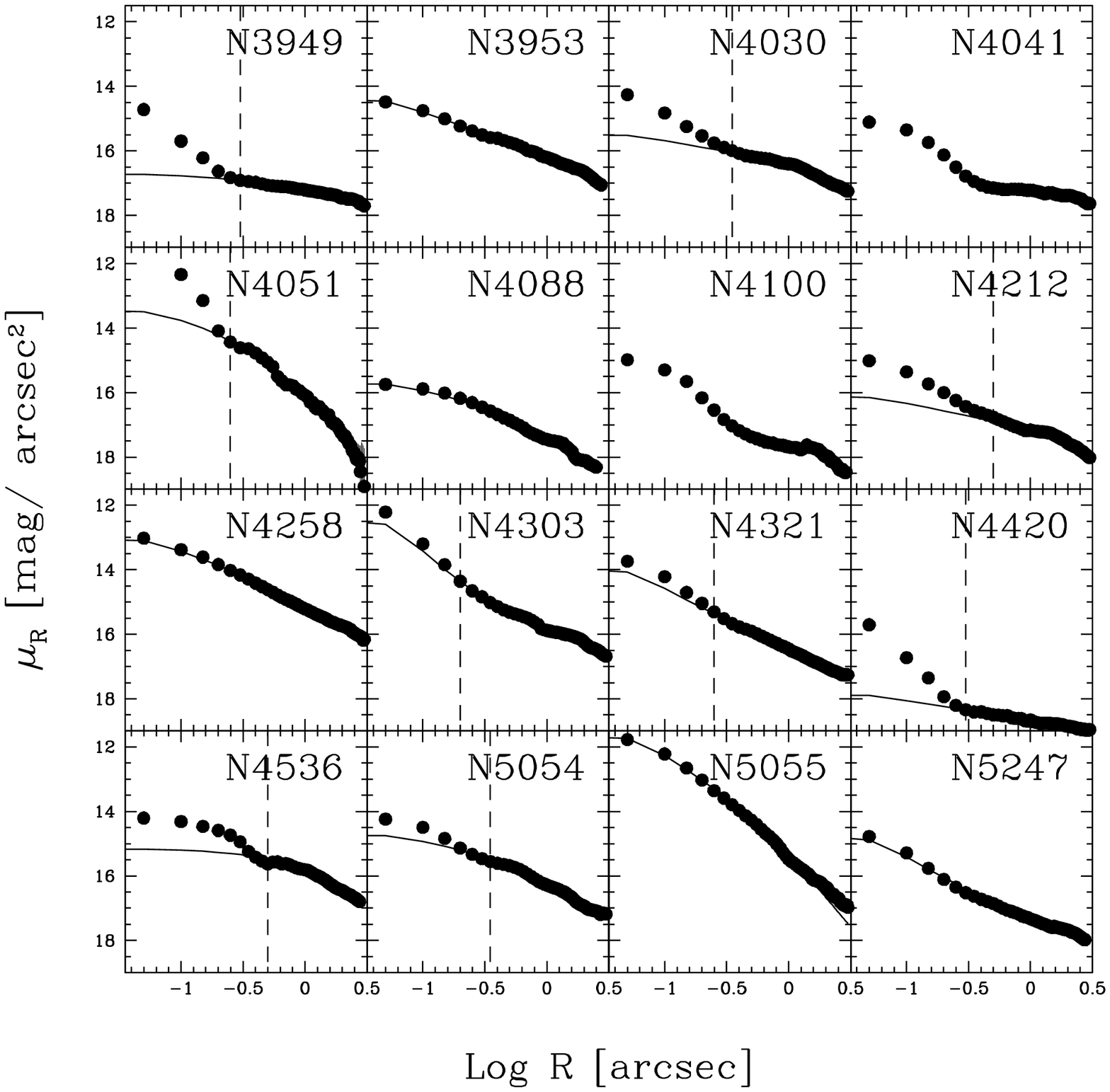}
\caption{Continued}
\end{figure}
\begin{figure}
\figurenum{2}
\plotone{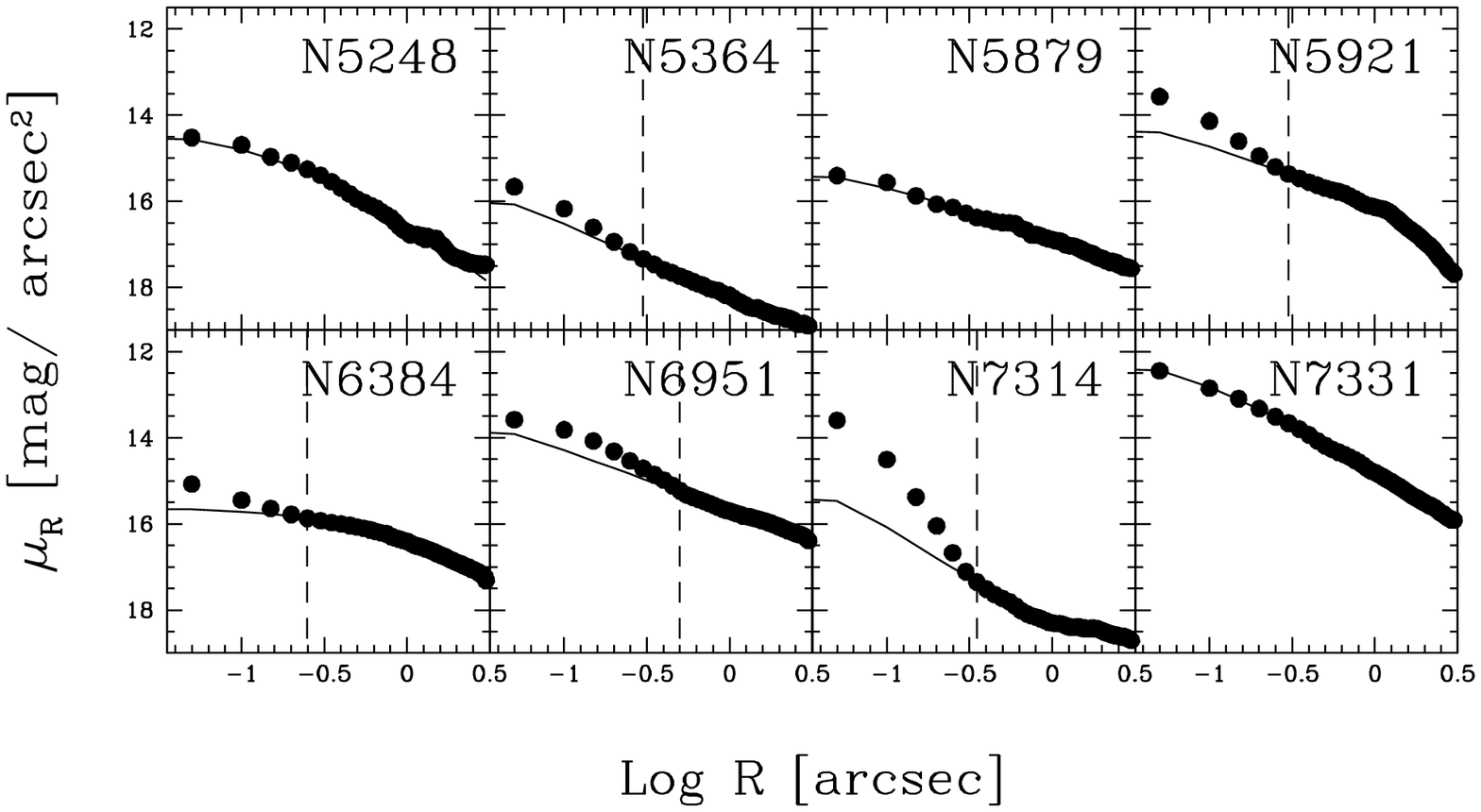}
\caption{Continued}
\end{figure}
\begin{figure}
\plotone{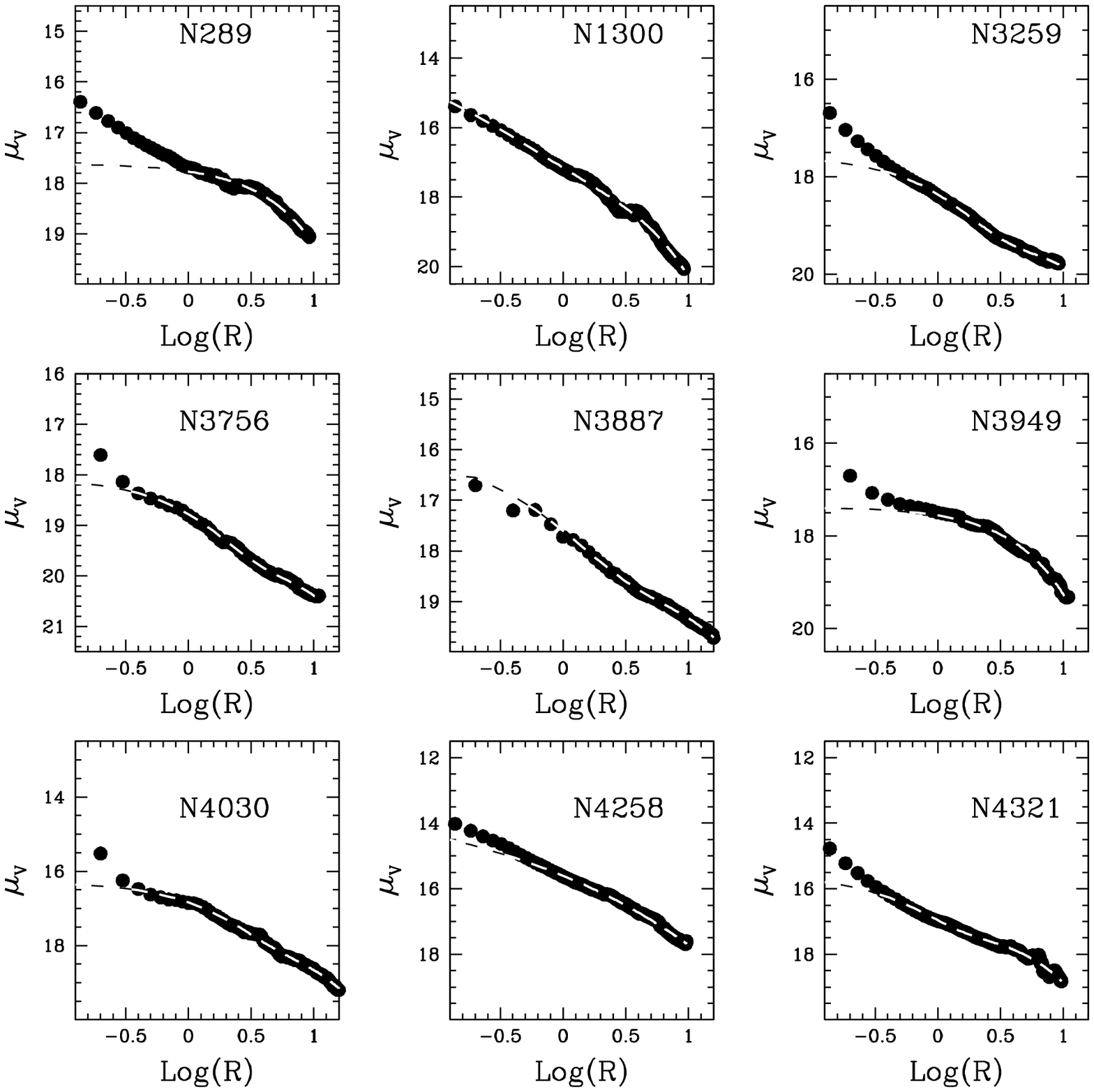}
\caption{$V-$band surface brightness profiles extracted from the WFPC2 data ({\it dots}) and best--fit model from column (8) of Table~\ref{tab:results} ({\it dashed lines}). The model was convolved with the WFPC2 PSF before plotting.
\label{fig:wfpcprof}}
\end{figure}
\begin{figure}
\figurenum{3}
\plotone{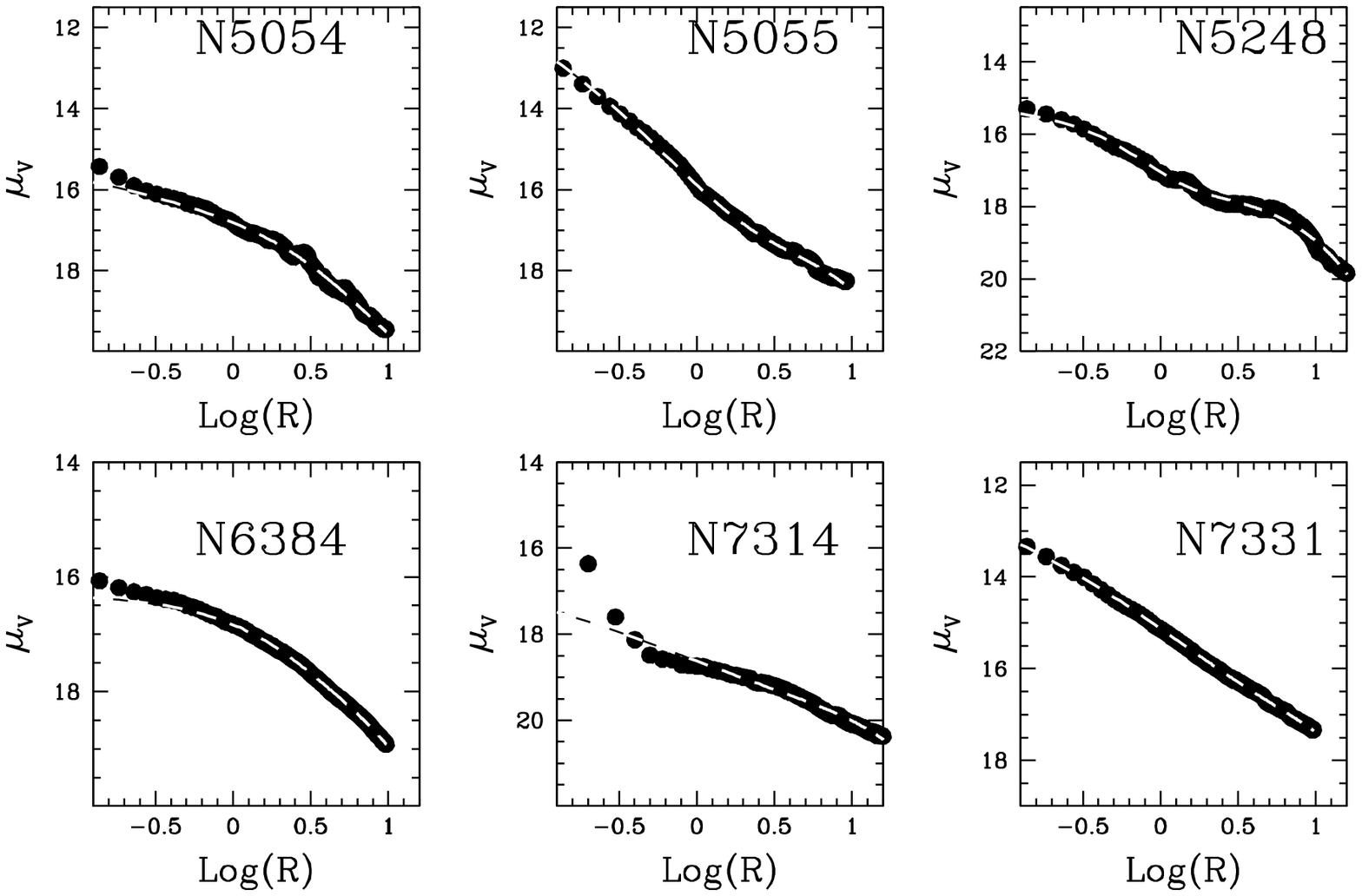}
\caption{Continued}
\end{figure}
\begin{figure}
\plotone{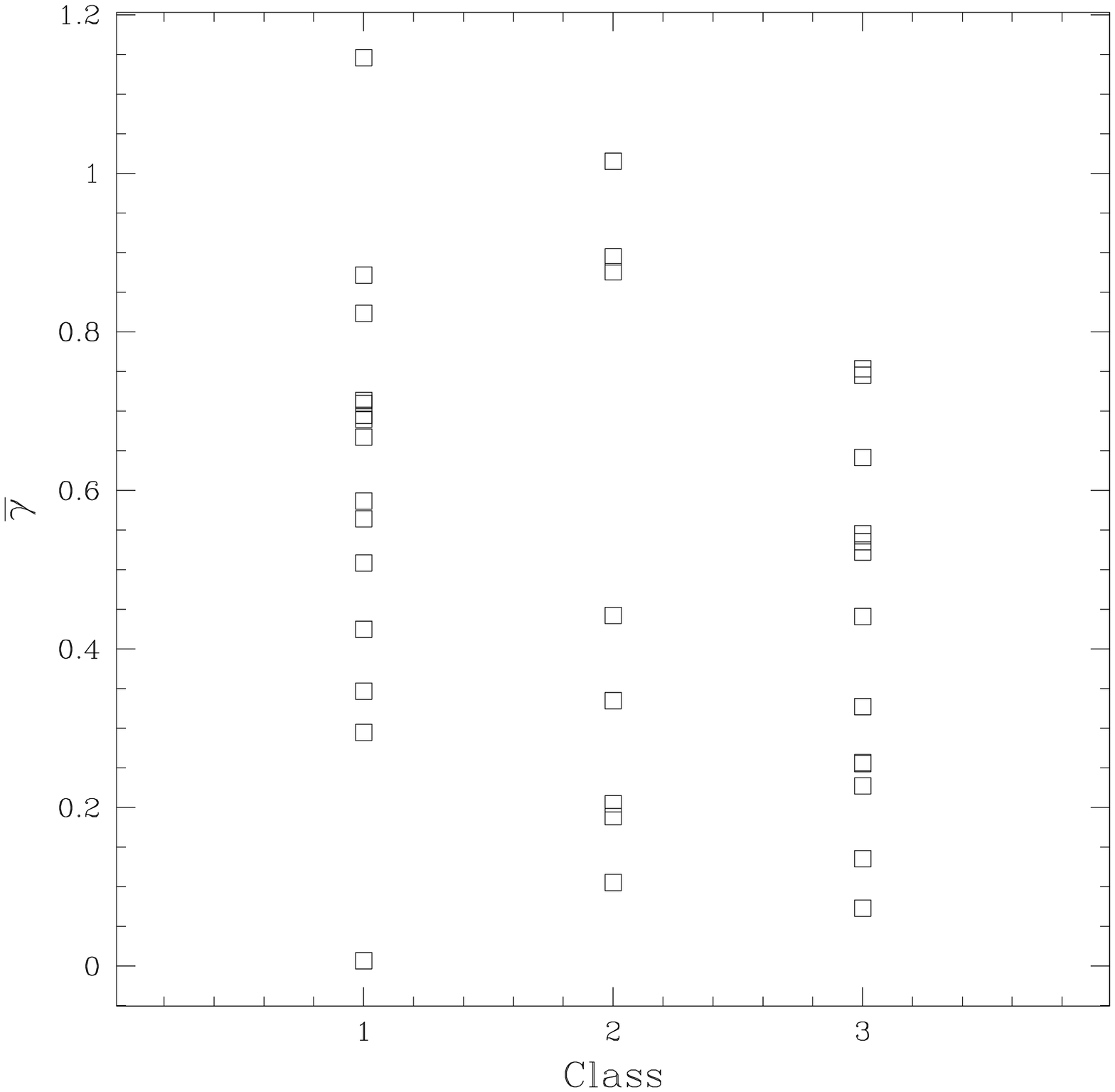}
\caption{Average logarithmic  central cusp slope  $\am$\  as a function of  the
class of the object:  class$=1$ means no central component;  class$=2$
means  point--like  central component; and   class$=3$ means  resolved
central component. This figure indicates that  there is no correlation
between  the presence  of  a nuclear  source and  the  value of $\am$\
measured from the model.
\label{fig:gamma_cluster}}
\end{figure}

\begin{figure}
\plotone{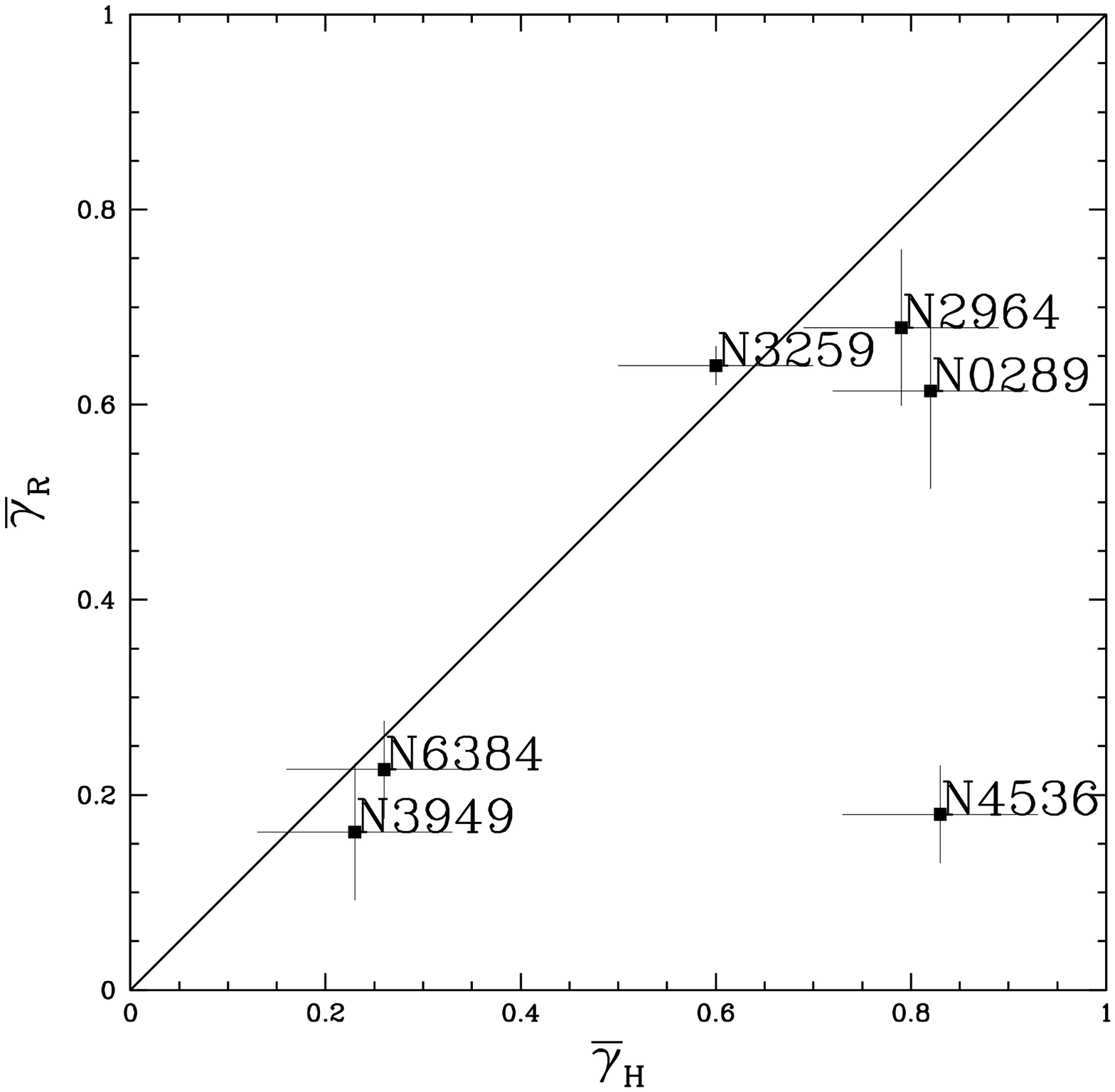}
\caption{Comparison between the average logarithmic slope ($\am$) 
measured by Seigar et al. (2002) ($\am_{H}$) from NICMOS $H-$band
images and the values computed in this paper ($\am_{R}$). The names of the
galaxies are used to label the points.  The galaxy for which $\am$\ is
most different (NGC~4536)  is strongly affected by dust in the STIS image (see
Figure~\ref{fig:images1}).
\label{fig:seigar}}
\end{figure}

\begin{figure}
\plotone{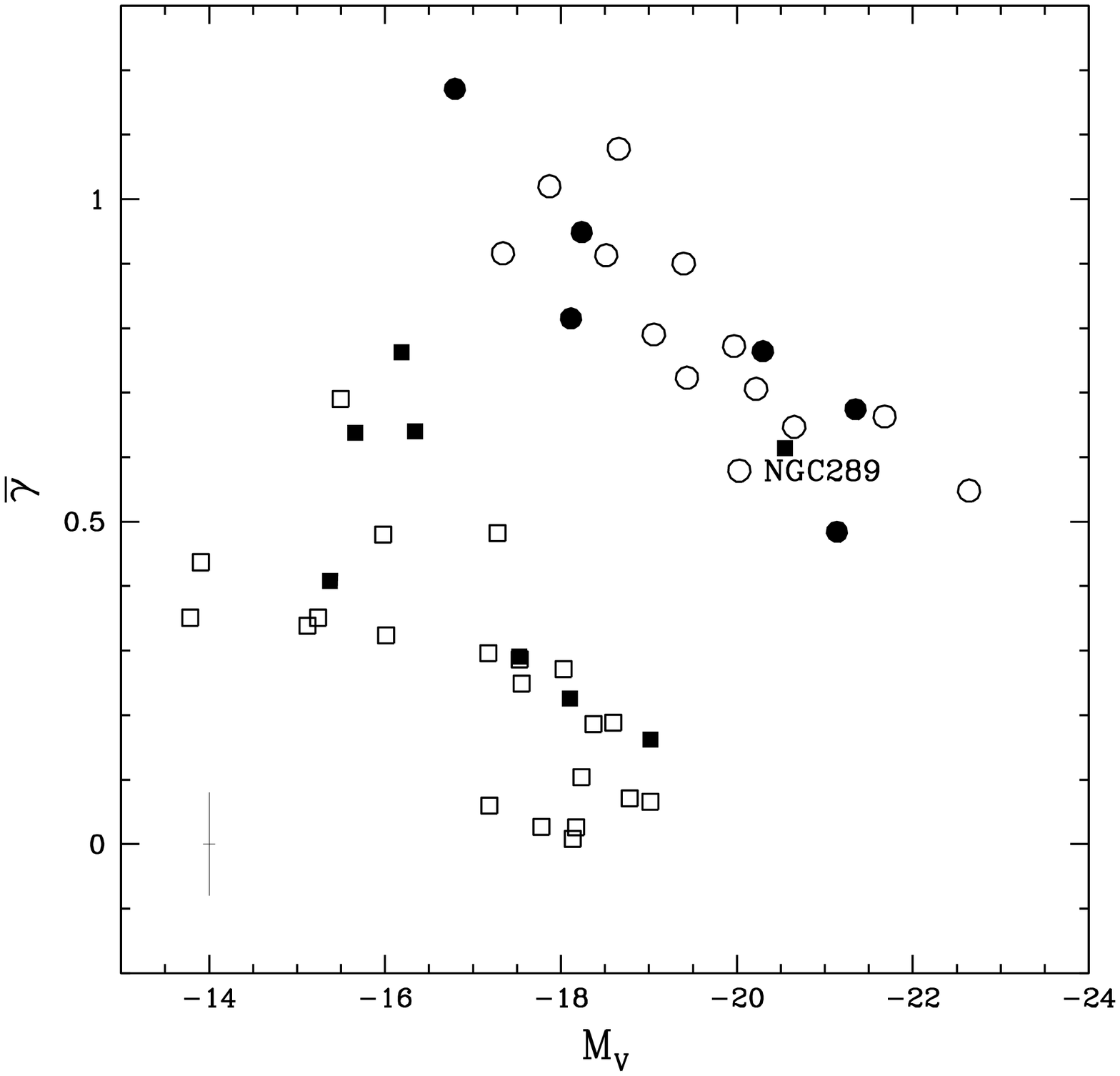}
\caption{Inner cusp slope $\am$\ versus $V-$band absolute magnitude of the bulge component. 
{\it Squares} and {\it circles} represent bulges with exponential
and $r^{1/4}$ SBPs, respectively. The galaxies from this study are the
{\it filled symbols} and those from Carollo \& Stiavelli (1998) are
the {\it open symbols}. From our sample only those galaxies are shown for
which archive WFPC2 data was available. \label{fig:gammabulge}}
\end{figure}

\begin{figure}
\plotone{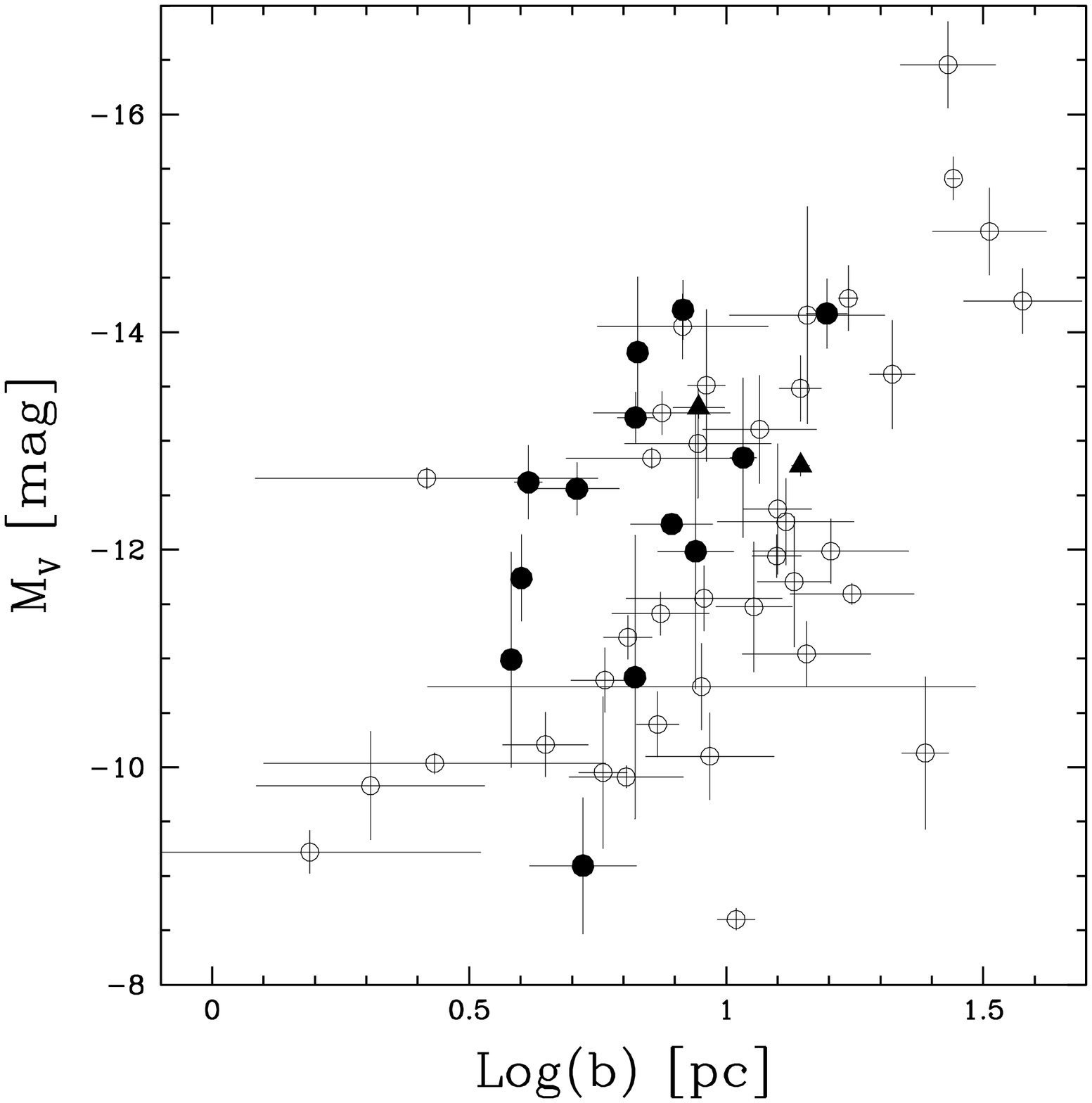}
\caption{Absolute $V$ magnitude and logarithm of the half light radius of the 
nuclear clusters identified in this work ({\it filled circles} and
{\it triangles}) and in Carollo \ea (2002, {\it open circles}). {\it
Triangles} represent galaxies for which we do not have the Nuker fit
of the bulge component. Our sample seems to have brighter magnitudes
on average for the same cluster size, especially for smaller $b$. This
could be due to the fact that almost all our clusters are
spectroscopically classified as star forming from ground based data
(\citealt{ho1}, see text for details).
\label{fig:clusters}}
\end{figure}

\clearpage
\begin{deluxetable}{lccllcc}
\tablecaption{Galaxy sample.\label{tab:sample}}
\tabletypesize{\small}
\tablecolumns{7}
\tablehead{
\multicolumn{1}{c}{Name} &
\multicolumn{1}{c}{Type} &
\multicolumn{1}{c}{$m_B$} &
\multicolumn{1}{c}{$A_R$} &
\multicolumn{1}{c}{$D$} &
\multicolumn{1}{c}{Scale} &
\multicolumn{1}{c}{Comments}\\
\multicolumn{1}{c}{}&
\multicolumn{1}{c}{}&
\multicolumn{1}{c}{(mag)}&
\multicolumn{1}{c}{(mag)}&
\multicolumn{1}{c}{Mpc}&
\multicolumn{1}{c}{pc/pixel}&
\multicolumn{1}{c}{} \\
\multicolumn{1}{c}{(1)}&
\multicolumn{1}{c}{(2)}&
\multicolumn{1}{c}{(3)}&
\multicolumn{1}{c}{(4)}&
\multicolumn{1}{c}{(5)}&
\multicolumn{1}{c}{(6)}&
\multicolumn{1}{c}{(7)}
}
\startdata 
{\objectname[]{NGC~134}}    & .SXS4../SBbc    & 11.25 &0.048        &   24.49   &    5.93  & missed\\
{\objectname[]{NGC~157}}    & .SXT4../SBbc    & 11.05 &0.119        &   27.64   &    6.70  & strong dust lane across the center\\
{\objectname[]{NGC~255}}    & .SXT4../SBbc    & 12.31 &0.071        &    26.30  &    6.37  & bright knots spread over the entire galaxy, \\
                            &                   &       &            &           &           & the brightest is not at the center of the \\
                            &                   &       &            &           &           & external isophotes\\ 
{\objectname[]{NGC~289}}    & .SBT4../SBbc    & 11.79 &0.052        &   25.32   &    6.14  & dust on the plane of the disk,\\
                            &                   &       &            &           &           & the SBP keeps rising until the HST resolution \\
{\objectname[]{NGC~613}}    & .SBT4../SBbc    & 10.99 &0.052        &   22.98   &    5.57  & disturbed morphology, \\
            &                   &       &            &           &           & dust lanes and bright knots. \\
            &                   &       &            &           &           & Nuclear dust ring at 0\farcs25 from the nucleus\\

{\objectname[]{NGC~1255}}    & .SXT4.. /SBbc     & 11.68 &0.037     &   25.49   &    6.18  & regular morphology \\
{\objectname[]{NGC~1300}}    & .SBT4../SBbc     & 11.22 &0.081      &   24.10   &    5.84   & regular morphology, spiral--like dust lanes \\
            &                   &       &            &           &           & down to the center \\
{\objectname[]{NGC~1832}}    & .SBR4../SBbc     & 12.12 &0.195      &   28.63   &   6.94    & spiral arms down to the center, \\
            &                   &       &            &           &           & wrapping around the nucleus\\
{\objectname[]{NGC~2748}}    & .SA.4../Sbc      & 12.41 &0.071      &   25.68   &   6.22   & asymmetric morphology due to diffuse dust\\
            &                   &       &            &           &           & dust lanes, nucleus not well defined \\
{\objectname[]{NGC~2903}}    & .SXT4../SBbc     &  9.56 &0.083      &   7.34    &   1.78    & spiral arms down to the center,\\
            &                   &       &            &           &           & many star--forming bright knots. \\
{\objectname[]{NGC~2964}}    & .SXR4*./SBbc     & 12.26 &0.052      &   19.92   &    4.83  & dust lanes and bright knots  \\
{\objectname[]{NGC~3003}}    & .S..4\$./SBbc     & 12.41 &0.036     &   22.47   &    5.44  & a clear center is not present, bright knots  \\
                           
{\objectname[]{NGC~3021}}    & .SAT4*./Sbc      & 12.60 &0.036      &   23.42   &    5.67  & dusty spiral arms down to the center, \\
            &                  &       &            &           &          & bright central nucleus, possiby a nuclear bar\\
{\objectname[]{NGC~3162}}    & .SXT4../SBc      & 12.38 &0.062      &   18.77   &    4.55  & wrapped spiral arm, or nuclear ring, \\
{\objectname[]{NGC~3254}}    & .SAS4../Sbc      & 12.62 &0.039        & 20.08   &    4.87  & narrow dust lane close to the nucleus, \\
            &                   &       &             &           &           & smooth underlying surface brightness \\
{\objectname[]{NGC~3259}}    & .SXT4*./SBbc     & 12.90 &0.039      &   27.98   &    6.78  & smooth surface brightness \\
{\objectname[]{NGC~3310}}    & .SXR4P./SBbc     & 11.08 &0.060      &   16.48   &    3.99  & strong dust lane crossing the center of the galaxy\\
{\objectname[]{NGC~3403}}    & .SA.4*./Sbc      & 13.08 &0.277        &   22.08 &    5.35  & flocculent spiral arms, knots of star formation, \\
            &                   &       &             &           &           & resolved component \\
{\objectname[]{NGC~3521}}    & .SXT4../SBbc     &  9.73 &0.155      &    9.46   &    2.29  & missed\\
{\objectname[]{NGC~3642}}    & .SAR4*./Sbc      & 11.67 &0.029      &   26.10   &    6.32  & dusty spiral arms, elongated nuclear structure, \\
            &                   &       &             &           &           & probably due to a dust lane close to the center\\
{\objectname[]{NGC~3684}}    & .SAT4../Sbc      & 12.27 &0.069        &   16.20 &    3.92  & disturbed morphology, strong dust lanes  \\
            &                   &       &             &           &           & many bright knots of star formation \\
{\objectname[]{NGC~3686}}    & .SBS4../SBbc     & 11.96 &0.065      &   16.10   &    3.90  & bright central component, resolved\\
{\objectname[]{NGC~3756}}    & .SXT4../SBbc     & 12.33 &0.031        & 21.17   &    5.13  & dusty spiral arms \\
{\objectname[]{NGC~3887}}    & .SBR4../SBbc     & 11.42 &0.092      &   14.79   &    3.58  & dust lanes definig tightly wrapped spiral arms,\\
            &                   &       &             &           &           & hints of a nuclear bar \\
{\objectname[]{NGC~3949}}    & .SAS4*./Sbc      & 11.70 &0.057      &   13.17   &     3.19  & dust lanes \\
{\objectname[]{NGC~3953}}    & .SBR4../SBbc     & 10.85 &0.080      &   17.41   &     4.22  & dust lane passing close to the nucleus\\
{\objectname[]{NGC~3972}}    & .SAS4../SBbc     & 12.99 &0.037      &   14.45   &     3.50  & missed \\
{\objectname[]{NGC~4030}}    & .SAS4../Sbc      & 11.67 &0.071      &   19.61   &     4.75  & flocculent spiral arms \\
{\objectname[]{NGC~4041}}    & .SAT4*./Sbc      & 11.85 &0.047      &   20.85   &    5.05  & flocculent spiral arms down to the center\\
{\objectname[]{NGC~4051}}    & .SXT4../SBbc     & 10.90 &0.035      &   11.53   &    2.79  & very bright central component \\
{\objectname[]{NGC~4088}}    & .SXT4../SBc      & 11.16 &0.053      &   12.74   &    3.09  & flocculent spiral arms \\
            &                   &        &          &           &           & nucleus obscured by dust\\
{\objectname[]{NGC~4100}}    & PSAT4../Sbc      & 11.89 &0.062        & 17.55   &    4.25  & asymmetric  morphology, strong dust obscuration \\
            &                   &        &            &           &           & numerous knots of starformation \\
{\objectname[]{NGC~4212}}    & .SA.5*./Sc       & 11.88 &0.089      &17.50$^{a}$&    4.25   & dust absorption down almost to the center\\
{\objectname[]{NGC~4258}}    & .SXS4../SBbc     &  9.09 &0.043      &    7.78   &    1.89  & regular morphology, no visible spiral arms\\
{\objectname[]{NGC~4303}}    & .SXT4../SBbc     & 10.18 &0.060      &   21.81   &    5.28  & small flocculent spirals arms and filaments\\
{\objectname[]{NGC~4321}}    & .SXS4../SBbc     & 10.07 &0.070        & 22.69   &    5.50  & dust absorption on the disk plane\\
            &                        &       &             &           &           & no well defined spiral arms \\
{\objectname[]{NGC~4389}}    & .SBT4P*/SBbc      & 12.59 &0.039     &   11.82   &    2.86  & missed  \\
{\objectname[]{NGC~4420}}    & .SBR4*./SBc       & 12.86 &0.048     &   23.45   &    5.68  & no well defined dust lanes\\
{\objectname[]{NGC~4527}}    & .SXS4../SBbc      & 11.45 &0.059     &   24.16   &    5.85  & severe dust obscuration, nucleus not well defined \\
{\objectname[]{NGC~4536}}    & .SXT4../SBbc      & 11.15 &0.049     &   25.24   &    6.11  & asymmetric structure due to dust obscuration \\
{\objectname[]{NGC~5005}}    & .SXT4../SBbc      & 10.74 &0.038     &   14.88   &    3.60  & nucleus not well defined, dust obscuration \\
{\objectname[]{NGC~5054}}    & .SAS4../Sbc       & 11.84 &0.220     &   23.44   &    5.68  & flocculent and not well defined spiral arms,\\
{\objectname[]{NGC~5055}}    & .SAT4../Sbc       &  9.30 &0.047     &   8.447   &    2.05  & no visible spiral arms or dust lanes \\
{\objectname[]{NGC~5247}}    & .SAS4../SBbc      & 11.17 &0.237     &   17.63   &    4.27  & patchy dust absorption \\
{\objectname[]{NGC~5248}}    & .SXT4../SBbc      & 11.01 &0.065     &   16.12   &    3.90  & well defined starforming spiral arm\\
{\objectname[]{NGC~5364}}    & .SAT4P./Sbc       & 11.37 &0.073     &   17.36   &    4.20  & regular morphology, resolved central cluster \\
{\objectname[]{NGC~5577}}    & .SAT4*./Sbc       & 13.74 &0.109     &   21.29   &    5.16  & missed \\
{\objectname[]{NGC~5713}}   & .SXT4P./SBbc      & 11.87 &0.105      &   27.40   &    6.64  & missed\\
{\objectname[]{NGC~5879}}    & .SAT4*\$/Sbc       & 12.18 &0.033            &   14.29   &    3.46  & dust lane down to very close to the center \\
                           
{\objectname[]{NGC~5921}}    & .SBR4../SBbc      & 11.76 &0.107        &        21.95   &    5.32  & regular morphology\\
            &                    &       &             &           &           & several spiral arms defined by dust lanes \\
{\objectname[]{NGC~6384}}    & .SXR4../SBbc      & 11.61 &0.330     &   26.49   &    6.42  & regular morphology, weak dust lane \\
{\objectname[]{NGC~6951}}    & .SXT4../SBbc      & 11.99 &0.978        &        26.30   &    6.37  & spiral arms down to the center \\
{\objectname[]{NGC~7314}}    & .SXT4../SBbc      & 11.68 &0.057     &   22.92   &    5.55  & very bright nucleus, smooth surface brightness\\
{\objectname[]{NGC~7331}}    & .SAS3../Sbc       & 10.26 &0.242     &   17.16   &    4.16  & regular morphology, no visible spiral arms \\
\enddata 
\tablecomments{
Col.(1): Galaxy name.  Col.(2): Morphological classification from RC3
\citep[][{\it left}]{devaucouleurs1} and PGC \citep[][{\it
right}]{paturel2003}.  Col.(3): Total observed blue magnitude from
RC3.  Col.(4): $R$ band galactic extinction from Schlegel \ea (1988).
Col.(5): Galaxy distance in Mpc.  Col.(6): Image scale in parsec per
STIS pixel.  Col.(7): Description of the nuclear morphology of the
galaxy.}
\end{deluxetable} 

\clearpage
\begin{table*}
\caption{Results of surface brightness profile fits.} 
\label{tab:results}
{\small\begin{center}
\begin{tabular}{lcccccccccc}
\hline\hline
\multicolumn{1}{l}{Name} &
\multicolumn{1}{c}{$\am$} &
\multicolumn{1}{c}{$\alpha$} &
\multicolumn{1}{c}{$\beta$} &
\multicolumn{1}{c}{$\gamma$} &
\multicolumn{1}{c}{$r_b$} &
\multicolumn{1}{c}{$I_b$ ($R-$band)} &
\multicolumn{1}{c}{Bulge} &
\multicolumn{1}{c}{$r_{0,{\rm bulge}}$} &
\multicolumn{1}{c}{m$_{V,{\rm bulge}}$} &
\multicolumn{1}{c}{Sersic $n$} \\
\multicolumn{1}{l}{} &
\multicolumn{1}{c}{} &
\multicolumn{1}{c}{} &
\multicolumn{1}{c}{} &
\multicolumn{1}{c}{} &
\multicolumn{1}{c}{[arcsec]} &
\multicolumn{1}{c}{[mag/arcsec$^2$]} &
\multicolumn{1}{c}{($r>1''$)} &
\multicolumn{1}{c}{[arcsec]} &
\multicolumn{1}{c}{[mag]} &
\multicolumn{1}{c}{($r>1''$)} \\
\multicolumn{1}{c}{(1)} &  
\multicolumn{1}{c}{(2)} &  
\multicolumn{1}{c}{(3)} &  
\multicolumn{1}{c}{(4)} &  
\multicolumn{1}{c}{(5)} &  
\multicolumn{1}{c}{(6)} &  
\multicolumn{1}{c}{(7)} &  
\multicolumn{1}{c}{(8)} &  
\multicolumn{1}{c}{(9)} &
\multicolumn{1}{c}{(10)} &  
\multicolumn{1}{c}{(11)} \\  

\hline
\hline \small

NGC~289  &0.61&   0.00&   0.72 &  0.56 &  1.09 & 17.17  & expo  & 6.73 & 11.47  & 1.33\\  
NGC~1255 &0.39&  17.87&   0.46 &  0.41 &  1.26 & 18.48  & ...           & ...   & ...   & ... \\  
NGC~1300 &0.81&  37.59&   1.50 &  0.85 &  1.91 & 17.40  &$r^{1/4}$      & 3.94 & 14.30 & 0.34\\  
NGC~1832 &0.35&   1.68&   3.15 &  0.29 &  2.36 & 17.40  & ...           & ...  & ...     & ...\\  
NGC~2964 &0.68&   1.42&   1.94 &  0.40 &  0.68 & 16.35  & NGF   & ...  & ...            & ... \\  
NGC~3021 &0.82&   0.01&   0.80 &  0.91 &  0.93 & 17.20  & ...   & ...  &  ...           &  ...\\  
NGC~3162 &0.30&   5.92&   1.27 &  0.22 &  0.47 & 16.77  & ...   & ...  & ...            &  ...\\  
NGC~3254 &0.33&   0.71&   1.35 &  0.00 &  1.17 & 16.88  &  ...  & ...  & ...            & ... \\  
NGC~3259 &0.64&   0.58&   0.67 &  0.67 &  1.08 & 18.17  &expo   &0.97  &15.89   & 0.81\\  
NGC~3403 &0.15&   0.63&   0.18 &  0.15 &  2.18 & 18.87  & ...   & ...  & ...            & ... \\  
NGC~3642 &0.89&   0.92&   1.58 &  0.60 &  0.53 & 16.06  &  ...  & ...  & ...            & ... \\  
NGC~3684 &0.09&  11.17&   0.59 &  0.10 &  0.74 & 17.62  & ...   & ...  &  ...           & ... \\  
NGC~3686 &0.86&   0.38&   0.55 &  1.14 &  0.60 & 17.29  & ...   & ...  & ...            & ... \\  
NGC~3756 &0.41&  11.22&   0.89 &  0.43 &  1.11 & 18.51  & expo          & 1.04  &16.26  & 0.90\\  
NGC~3887 &0.45&   1.91&   0.97 &  0.33 &  0.53 & 16.72  & \ser          & 0.49 &16.48   & 0.55\\  
NGC~3949 &0.16&   0.87&   0.41 &  0.00 &  0.38 & 16.95  & expo          & 5.74 &11.58   & 0.72\\  
NGC~3953 &0.57&   8.02&   3.29 &  0.59 &  2.75 & 17.10  & ...           & ...  & ...    & ... \\  
NGC~4030 &0.29& 145.71&   0.79 &  0.30 &  1.15 & 16.44  & expo  & 2.00 &13.94   & 0.96\\  
NGC~4051 &0.82&   0.61&   3.65 &  0.05 &  1.90 & 17.26  & ...           & ...  & ...    & ... \\  
NGC~4088 &0.54&   0.51&   1.78 &  0.00 &  1.08 & 17.48  &  ...          & ...  & ...    & ... \\  
NGC~4212 &0.33&  40.54&   1.00 &  0.35 &  1.64 & 17.34  & ...           & ...  & ...    & ... \\  
NGC~4258 &0.67&   0.27&   1.64 &  0.00 &  0.70 & 14.84  & $r^{1/4}$     &87.56  & 8.11  & 0.33\\  
NGC~4303 &1.06&   8.21&   0.65 &  1.28 &  0.32 & 14.95  & ...           & ...  & ...    & ... \\  
NGC~4321 &0.76&   0.14&   0.02 &  1.39 &  1.73 & 16.90  & expo  & 0.45 &15.59   & 0.95\\  
NGC~4420 &0.26&   0.11&   0.01 &  0.45 &  1.38 & 18.76  & ...           & ...  & ...    & ... \\  
NGC~4536 &0.18&   1.09&   2.19 &  0.00 &  2.34 & 16.62  & NGF           &  ... & ...    & ... \\  
NGC~5054 &0.48&   0.52&   1.68 &  0.03 &  1.40 & 16.54  & $r^{1/4}$  & 30.54 &10.71  & 0.30\\
NGC~5055 &1.17&   0.37&   2.96 &  0.00 &  0.63 & 14.62  & $r^{1/4}$  & 1.00 &12.84      & 0.27\\  
NGC~5247 &0.85&  18.05&   0.54 &  0.89 &  0.52 & 16.92  &  ...          &  ... &  ...   &  ...\\  
NGC~5248 &0.64&   0.48&   1.92 &  0.00 &  0.92 & 16.43  & expo          &0.38  &15.37   & 1.02\\  
NGC~5364 &0.70&   1.35&   0.33 &  0.77 &  1.88 & 18.63  & ...           & ...  & ...    & ... \\  
NGC~5879 &0.48&   0.14&   1.06 &  0.00 &  0.58 & 16.57  & ...           & ...  & ...    &  ...\\  
NGC~5921 &0.56&   8.15&   1.62 &  0.58 &  1.41 & 16.43  &  ...          & ...  & ...    &  ...\\  
NGC~6384 &0.23&   0.79&   1.60 &  0.00 &  2.36 & 17.02  & expo          &1.38  & 14.01  & 0.84\\  
NGC~6951 &0.58&   0.32&   0.28 &  0.79 &  1.31 & 15.84  & NGF   & ...  &  ...           & ... \\  
NGC~7314 &0.95&  17.40&   0.31 &  0.99 &  0.77 & 18.16  & \dvc  & 22.33 & 13.56 & 0.24\\  
NGC~7331 &0.76&   0.21&   1.86 &  0.00 &  0.98 & 14.76  & $r^{1/4}$  &7.62  &10.88      & 0.23\\  
\hline
\end{tabular}

\begin{minipage}{18cm}  
Note. -- Col.(2): average logarithmic slope between 0\farcs1 and
0\farcs5. Col.(3--7): Nuker best--fit parameters ($\alpha$, $\beta$,
$\gamma$; $r_b$ in arcsec, and $I_b$ in mag/arcsec$^2$). The fit was
done to the STIS data and excluded the very central pixels if a
nuclear component is present.  Col.(8): bulge classification derived
from the WFPC2 surface brightness profile for $r>1''$, NGF indicates
that no meaningful isophotal fit could be extracted for the galaxy.
Col.(9): bulge scale radius in arcsec defined as the effective radius
of the \dvc\ model or the scale radius for the exponential bulges.
Col.(10): total apparent $V$ magnitude of the bulge component.
Col.(11): index $n$ of the best--fitting Sersic model.
\end{minipage}  
\end{center}
}
\end{table*}
\normalsize
        
\clearpage      
\begin{table*}
\caption{Properties of central components.}
\label{tab:clusters}
{\begin{center}\begin{tabular}{lccccc}
\hline\hline
\multicolumn{1}{l}{Galaxy} &
\multicolumn{1}{c}{Classification} &
\multicolumn{1}{c}{M$_{R}$} &
\multicolumn{1}{c}{$b$} &
\multicolumn{1}{c}{FWHM} &
\multicolumn{1}{c}{Spectrum} \\
\multicolumn{1}{l}{} &
\multicolumn{1}{c}{} &
\multicolumn{1}{c}{[mag]} &
\multicolumn{1}{c}{[pc]} &
\multicolumn{1}{c}{['']} &
\multicolumn{1}{c}{} \\
\multicolumn{1}{l}{(1)} &
\multicolumn{1}{c}{(2)} &
\multicolumn{1}{c}{(3)} &
\multicolumn{1}{c}{(4)} &
\multicolumn{1}{c}{(5)} &
\multicolumn{1}{c}{(6)} \\
\hline
\hline 
NGC~1255 &  PS  &      ...           &   ...        &     ...   &   ... \\
NGC~3021 &  R   &  -14.67$\pm 0.3$   & 7.7$\pm  0.1$&0.16       &  ...  \\
NGC~3162 &  R   &  -13.08$\pm 0.6$   & 4.1$\pm  0.1$&0.13       & HII   \\
NGC~3259 &  PS  &     ...            &   ...       &  ...       &  ...  \\
NGC~3403 &  R   &  -9.56$\pm 0.3$    & 5.1$\pm  0.5$&0.13       &  ...  \\
NGC~3684 &  PS  &  ...               &   ...        &    ...    & HII   \\      
NGC~3686 &  PS  &   ...              &   ...        &   ...     & HII   \\      
NGC~3756 &  R   &  -12.10$\pm 0.6$   & 4.0$\pm  0.1$&0.12       & HII   \\
NGC~3887 &  R   &  -11.45$\pm 0.6$   & 3.8$\pm  0.1$&0.14       &  ...  \\
NGC~3949 &  PS  &      ...           &   ...        &    ...    & HII   \\
NGC~3953 &  PS  &      ...           &   ...        &    ...    & T2    \\      
NGC~4030 &  R   &  -13.68$\pm 0.3$   & 6.6$\pm  0.2$&0.16       &  ...  \\
NGC~4041 &  R   &  -13.24$\pm 0.1$   &14.2$\pm  0.2$&0.25       & HII   \\
NGC~4051 &  PS  &      ...           &   ...        &    ...    & S1.2  \\      
NGC~4100 &  R   &  -12.70$\pm 0.1$   & 5.1$\pm  0.6$&0.15       & HII   \\
NGC~4212 &  R   &  -13.02$\pm 0.2$   & 7.8$\pm  0.4$&0.19       &  ...  \\      
NGC~4303 &  PS  &      ...                   &   ...        &    ...    & HII   \\      
NGC~4321 &  R   &  -13.77$\pm 0.1$   & 8.8$\pm  0.4$&0.17       & T2    \\
NGC~4420 &  PS  &        ...         &   ...        &    ...    &   ... \\      
NGC~4536 &  PS  &         ...        &   ...        &    ...    & HII   \\      
NGC~5054 &  R   &  -13.31$\pm 1.3$   &10.8$\pm  0.3$&0.19       &  ...  \\
NGC~5364 &  R   &  -11.29$\pm 1.4$   & 6.6$\pm  0.1$&0.17       & HII   \\
NGC~5921 &  R   &  -14.28$\pm 0.8$   & 6.7$\pm  0.1$&0.15       & T2    \\
NGC~6384 &  R   &  -12.45$\pm 1.5$   &  8.7$\pm 0.6$&0.16       & T2    \\
NGC~6951 &  R   &  -14.63$\pm 0.1$   & 15.8$\pm 0.6$&0.23       & S2    \\
NGC~7314 &  PS  &     ...            &   ...        &   ...     & S1.9   \\     

\hline
\end{tabular}
\begin{minipage}{18cm}  
Note. -- Col.(2): Central component classification: Point Source (PS)
or resolved (R).  Col.(3): Absolute $R$ magnitude for resolved
sources.  Col.(4): Half light radius $b$ in parsec for resolved
sources.  Col.(5): Full Width at Half Maximum (in arcsec) of the
Gaussian used to derive $b$.  Col.(6): Spectral classification of the
nucleus from \citealt{ho1}; HII= star--forming nucleus, S= Seyfert
nucleus, and T= transition object. The number attached to the class
letter designates the type), except for NGC~7314 for which the
classification is taken from the NASA Extragalactic
Database\footnote{The NASA/IPAC Extragalactic Database (NED) is
operated by the Jet Propulsion Laboratory, California Institute of
Technology, under contract with the National Aeronautics and Space
Administration}.
\end{minipage}  
\end{center}}
\end{table*}

\end{document}